\documentclass[letter]{aa}
\usepackage{graphicx}
\usepackage{txfonts}
\usepackage{natbib}

\begin{document}

\title{Wavefront error correction and Earth-like planet \\detection by Self-Coherent Camera in space}

\author{Rapha\"{e}l~Galicher\inst{1}\fnmsep\inst{2}, Pierre~Baudoz\inst{1}\fnmsep\inst{2}, G\'{e}rard~Rousset\inst{1}\fnmsep\inst{2}}
\institute{LESIA, Observatoire de Paris, CNRS and University Denis Diderot Paris 7. 5, place Jules~Janssen, 92195 Meudon, France.\\
  \email{raphael.galicher@obspm.fr, pierre.baudoz@obspm.fr,gerard.rousset@obspm.fr}
 \and
 Groupement d'Int\'{e}r\^{e}t Scientifique Partenariat Haute R\'{e}solution Angulaire Sol Espace~(PHASE) between ONERA, Observatoire de Paris, CNRS and University Denis Diderot Paris 7}

\abstract
    {In the context of exoplanet detection, the performance of coronagraphs is limited by wavefront errors.}
    {To remove efficiently the effects of these aberrations using a deformable mirror, the aberrations themselves must be measured in the science image to extremely high accuracy.}
    {The Self-Coherent Camera which is based on the principle of light incoherence between star and its environment can estimate these wavefront errors. This estimation is derived directly from the encoded speckles in the science image, avoiding differential errors due to beam separation and non common optics.}
    {Earth-like planet detection is modeled by numerical simulations with realistic assumptions for a space telescope.}
    {The Self-Coherent Camera is an attractive technique for future space telescopes. It is also one of the techniques under investigation for the E-ELT planet finder the so-called EPICS.}
    
    \date{Received ; accepted }
    
    \titlerunning{Wavefront etimation by Self-Coherent Camera}
    \authorrunning{R.~Galicher et al.}
    
    \maketitle
    
    \section{Introduction}
    Very high contrast imaging is mandatory for the direct detection of exoplanets, which are typically a factor of between~$10^7$ and~$10^{10}$ fainter than their host and often located within a fraction of an arcsecond of their star. First of all, coronagraphs are required to suppress the overwhelming flux of the star but they are limited by wavefront errors in the upstream beam, which creates residual speckles in the science image. Adaptive optics must be used to correct for the effect of most of these aberrations. Some remain uncorrected generating quasi-static residual speckles\,\citep{Cavarroc06}. Interferential techniques take advantage of the incoherence between companion and stellar lights to measure these wavefront errors in the science image to high accuracy\,\citep{Codona04,Guyon04}. In this Letter, we describe such a technique called a Self-Coherent Camera\,\citep{Baudoz06}. Residual speckles in the science image, also called interferential image hereafter, are spatially encoded by fringes so that we can derive an estimation of wavefront errors to be corrected by a Deformable Mirror~(DM). Since the number of DM actuators is finite, this correction leaves residual speckles. Thus, after reaching the DM limit correction, we apply an image post-processing algorithm\,\citep{Galicher07} to achieve Earth-like planet imaging. Hereafter, we detail the SCC principle. Then, we describe the wavefront error estimator that we use. Finally, we present expected performances from space.
    
    \section{Principle and aberration estimator}
    \label{sec : principle}
    \begin{figure}[!ht]
      \resizebox{\hsize}{!}{\includegraphics{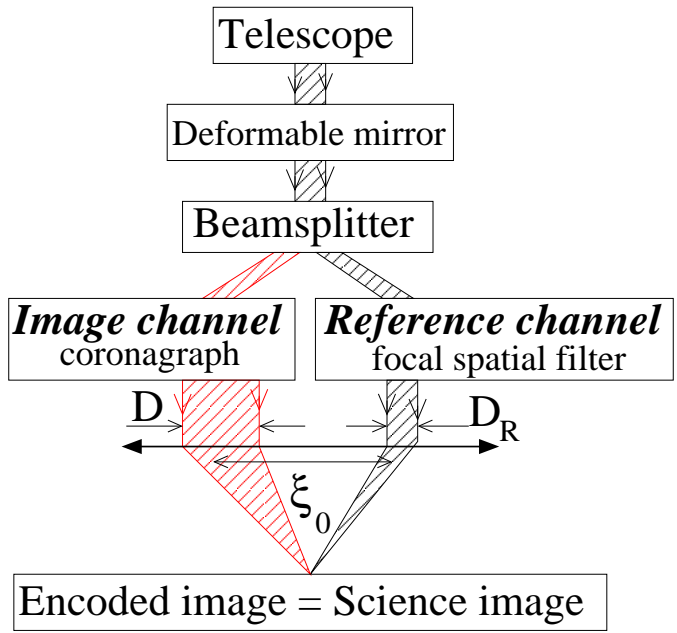}}
      \caption[SCC device schematics]{\it Self-Coherent Camera principle schematics.}
      \label{fig : schema}
    \end{figure}
\begin{figure}[!ht]
      {\includegraphics{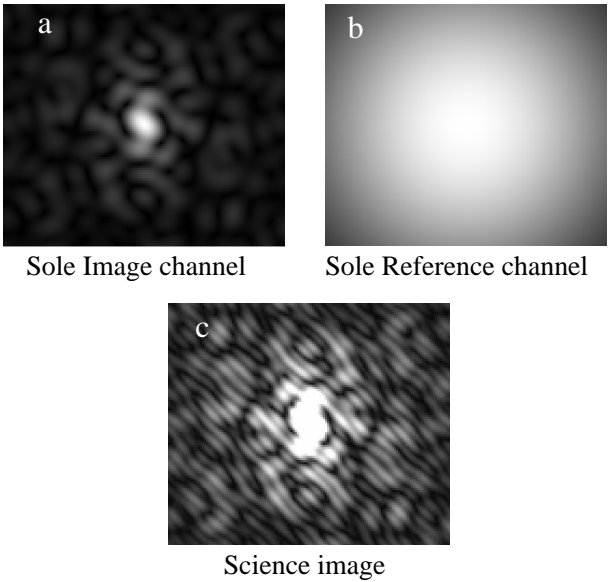}}
      \caption[Images in SCC]{\it (a)~Image formed after the sole coronagraph for a pupil of diameter~$D$~(sole image channel) showing the residual speckles. (b)~Image of the sole reference channel for a pupil of diameter~$D_R$. (c)~Interferential image~(science image) where the speckles are spatially encoded by fringes. The spatial scale is the same for all the images.}
      \label{fig : images}
    \end{figure}

    The beam from the telescope is reflected onto the DM and is split into two beams~(Fig.\,\ref{fig : schema}). The image channel~(shown in red in electronic edition) propagates through a coronagraph. It contains companion light and residual stellar light due to wavefront errors. Its complex amplitude is~$\Psi_{\mathrm{S}}(\xi) + \Psi_{\mathrm{C}}(\xi)$, where~$\xi$ is the pupil coordinate. $\Psi_{\mathrm{S}}$ and $\Psi_{\mathrm{C}}$ represent the stellar and companion complex amplitudes of the field in the pupil plane respectively, just after the $D$ diameter Lyot stop. The second beam, called the reference channel, is filtered spatially in a focal plane using a pinhole of radius smaller than $\lambda/D$. Almost all companion light is stopped since it is not centered on the pinhole. In the pupil plane just after the diaphragm~($D_{\mathrm{R}}$), the reference complex amplitude is called~$\Psi_{\mathrm{R}}(\xi)$. The pinhole reduces the impact of wavefront errors on $\Psi_{\mathrm{R}}$ since it acts as a spatial frequency filter. An optic recombines the two channels, separated by $\xi_0$ in the pupil plane, and creates a Fizeau fringed pattern in the focal plane. Residual speckles are therefore spatially encoded unlike companions. The mean intensity of residual speckles of the image channel is almost spatially flat and attenuated by the coronagraph. To optimize the fringe contrast, we have to match the intensity distributions and fluxes of image and reference channels. We use a~$D_{\mathrm{R}} < D$ diameter diaphragm to obtain an almost flat reference intensity in the focal plane. This diaphragm reduces the impact of aberrations on $\Psi_{\mathrm{R}}$, since only a fraction of the diffraction peak of size~$\lambda/D_{\mathrm{R}}$ is detected in the image~(image~b in Fig.~\,\ref{fig : images}). This implies that the reference channel is quite insensitive to aberrations and can be calibrated before the interference recording~\citep{Galicher07}. Fluxes are equalized using a variable neutral density in the reference channel before the pinhole~(Sect.\,\ref{sec : performances}). In Fig.\,\ref{fig : images}, we present, on the same spatial scale, (a)~the image formed after the sole coronagraph for a pupil of diameter~$D$~(the sole image channel) showing residual speckles, (b)~the image corresponding to the sole reference channel for a pupil of diameter~$D_R$ and (c)~the interferential image, where the residual speckles are spatially encoded by fringes.

In polychromatic light, the intensity~$I(\alpha)$ of the interferential image on the detector is
    \begin{eqnarray}
      & I(\alpha) = \int_{\mathcal{R}} \frac{1}{\lambda^2}\Big[I_{\mathrm{S}}\left(\frac{\alpha D}{\lambda}\right)+I_{\mathrm{R}}\left(\frac{\alpha D}{\lambda}\right) +I_{\mathrm{C}}\left(\frac{\alpha D}{\lambda}\right) \nonumber \\
        & + 2 Re \left( A_{\mathrm{S}}\left(\frac{\alpha D}{\lambda}\right)A_{\mathrm{R}}^*\left(\frac{\alpha D}{\lambda}\right) \exp{\left(\frac{2i\pi \alpha\xi_0}{\lambda}\right)}\right) \Big]d\lambda,
      \label{eq : intens_focal}
    \end{eqnarray}
    where~$\alpha$ is the angular coordinate in the science image, $A_i$ the Fourier tranform of the corresponding~$\Psi_i$, $I_i$ the intensity~$|A_i|^2$, and $A_i^*$ the conjugate of $A_i$. The wavelength $\lambda$ belongs \linebreak to ${\mathcal{R}}=[\lambda_0-\Delta\lambda/2,\lambda_0+\Delta\lambda/2]$. Following the work by \citet{Borde06}, we estimate wavefront errors from residual speckles in the science image. For this purpose, we propose to extract the modulated part of $I$, which contains a linear combination of $A_{\mathrm{S}}$ and $A_{\mathrm{R}}$. First, we apply a Fourier Transform on $I$ and isolate one of the lateral correlation peaks. We then apply an inverse Fourier Transform and obtain~$I_-$
    \begin{equation}
       I_-(\alpha) = \int_{\mathcal{R}}\frac{1}{\lambda^2}\, A_{\mathrm{S}}\left( \frac{\alpha\, D}{\lambda}\right)\,A_{\mathrm{R}}^*\left(\frac{\alpha\, D}{\lambda}\right)\, \exp{\left(\frac{2\,i\,\pi\, \alpha\, \xi_0}{\lambda}\right)}\,d\lambda
      \label{eq : I-_0}
    \end{equation}
    In Eq.~\ref{eq : I-_0}, $A_{\mathrm{S}}$ and $A_{\mathrm{R}}^*$ depend on $\alpha\, D/\lambda$, inducing the speckle dispersion with wavelength. Fizeau interfringe $\lambda/\xi_0$ is proportional to wavelength in the exponential term. Both effects degrade the wavefront estimation from $I_-$ when the useful bandwidth is large. However, the fringe wavelength dependence is dominant. It may be more appropriate to consider an Integral Field Spectrometer at modest resolution~($R=100$), or to simulate the use of a short bandpass filter with a chromatic compensator. Such a device, proposed by \citet{Wynne79}, almost correct for the two chromatic effects over a wide spectral band~($\Delta\lambda\simeq0.2\,\lambda_0$) to provide a smaller effective bandwidth~($\Delta\lambda_{\mathrm{eff}}\simeq 0.01\,\lambda_0$). It enables us to assume as close as possible a monochromatic case in our model of SCC image formation. We firstly assume $\Delta\lambda_{\mathrm{eff}}\ll\lambda_0$, so that $A_{\mathrm{S}}$ and $A_{\mathrm{R}}^*$ are constant over the spectral band. We obtain from Eq.~\ref{eq : I-_0}
    \begin{equation}
      I_-(\alpha) \simeq A_{\mathrm{S}}\left(\frac{\alpha\, D}{\lambda_0}\right)\,A_{\mathrm{R}}^*\left(\frac{\alpha\, D}{\lambda_0}\right)\, \int_{\mathcal{R}}\frac{1}{\lambda^2}\,\exp{\left(\frac{2\,i\,\pi\, \alpha\,\xi_0}{\lambda}\right)}\,d\lambda
      \label{eq : I-_1}
    \end{equation}
    We have to estimate $A_{\mathrm{S}}$ or, more precisely, its inverse Fourier transform $\Psi_{\mathrm{S}}$ and we deduce
    \begin{equation}
      \Psi_{\mathrm{S}}(\xi) \simeq \mathcal{F}^{-1}\left[\displaystyle \frac{I_-(\alpha)\,F^*(\alpha)}{A_{\mathrm{R}}^*(\alpha\, D/\lambda_0)\,\|F\|^2}\right],
      \label{eq : psiS_0}
    \end{equation}
    where $\mathcal{F}^{-1}$ denotes the inverse Fourier transform,
    \begin{equation}
      F = \displaystyle \int_{\mathcal{R}}\frac{1}{\lambda^2}\exp{(2\,i\,\pi\, \alpha\,\xi_0/\lambda)}d\lambda
    \end{equation}
    and $F^*$ its conjugate.

As a second assumption, we consider that wavefront errors $\phi$ we are attempting to measure are small and we can write the star field ${\Psi'}_{\mathrm{S}}$ in the pupil plane upstream from the coronagraph as
    \begin{equation}
      {\Psi'}_{\mathrm{S}}(\xi)\simeq\Psi_0\,P(\xi)\left(1+\frac{2\,i\,\pi\,\phi(\xi)}{\lambda_0}\right)
      \label{eq : psi_up}
    \end{equation}
    where $\Psi_0$ is the amplitude of the star assumed to be uniform over~$P$, which is the unitary flat pupil of diameter~$D$.

In a third step, we assume a perfectly achromatic coronagraph\,\citep{Cavarroc06}, which allows us to remove the coherent part of the energy $\Psi_0\,P$ to~${\Psi'}_{\mathrm{S}}$
    \begin{equation}
      \Psi_{\mathrm{S}}(\xi) \simeq \frac{2\,i\,\pi}{\lambda_0}\,\Psi_0\,P(\xi)\,\phi(\xi)
      \label{eq : psiS_1}
    \end{equation}
    Finally, Eq.~\ref{eq : psiS_0} and \ref{eq : psiS_1} provide an estimator of the wavefront errors within the pupil
    \begin{equation}
      \phi(\xi) \simeq  \frac{\lambda_0}{ 2\,\pi}\left[\mathcal{I}\left\{\mathcal{F}^{-1}\left(\frac{I_-(\alpha)\,F^*(\alpha)}{\Psi_0\, A_{\mathrm{R}}^*(\alpha\, D/\lambda_0)\, \|F\|^2}\right)\right\}\right]
      \label{eq : phase_est}
    \end{equation}
    with $\mathcal{I}\{\,\}$ the imaginary part. In Eq.~\ref{eq : phase_est}, $F$ depends only on known physical parameters, $\xi_0$ and the spectral bandwidth, and is numerically evaluated. We can estimate $\Psi_0$ since we can calibrate the incoming flux collected by the telescope. $I_-$ is derived from the recorded image~$I$. Finally, we have to divide by the complex amplitude $A_{\mathrm{R}}^*$, previously calibrated. Setting $D_{\mathrm{R}} << D$, we obtain an almost flat reference intensity and therefore avoid values close to zero in the numeric division. We notice in Eq.~\ref{eq : phase_est} the linear dependence of $\phi$ on $I_-$, which is measured directly from the interferential image. We attempt to correct for these wavefront errors, estimated from Eq.~\ref{eq : phase_est}, using the DM. Then, we record a new interferential image in which quasi-static residual speckles have been suppressed and companions are now detectable. Practically, few iterations are required to reach high contrast under our assumptions used to derive the estimator and because of noise.

    We note that we require SCC sampling sufficient to detect the fringes that encode the residual speckles. The sampling is then larger than the classical sampling used in earlier techniques proposed by \citet{Guyon04} and \citet{Codona04}. However, the SCC needs a single image to estimate wavefront errors, whereas the other two techniques require at least two images since they use an on-axis recombining as in a Michelson scheme and either temporal or spatial phase shifting arrangements. Finally, instead of spreading the incoming light into several images, the SCC spreads the light into fringes.
    
    \section{Performances}
    \label{sec : performances}
    We consider an SCC device operating in visible light~($\lambda_0\!=\!0.8\,\mu m$, $\Delta\lambda\simeq0.2\,\lambda_0$, $\Delta\lambda_{\mathrm{eff}}\!=\!0.01\,\lambda_0$, section\,\ref{sec : principle}). We assume a perfectly achromatic coronagraph. The beamsplitter injects $99\%$ of the incoming energy into the image channel. The filtering pinhole radius is $\lambda_0/D$ and $D$ equals $25\,D_{\mathrm{R}}$. To be more realistic, we assume a calibration of the reference channel with a non-aberrated incoming wavefront and enter this value into the expression for the estimator in Eq.~\ref{eq : phase_est}.  We consider a $32\times32$ DM. The nth-actuator influence function is~$\exp{(-1.22\,(32\,(\xi-\xi_{\mathrm{n}})/D)^2)}$, where $\xi_{\mathrm{n}}$ is the center of the nth-actuator. We call $\mathcal{H}$ the~$(32\,\lambda_0/D)^2$ corrected area which is centered on-axis. We chose $\xi_0=1.05\,(1.5\,D +0.5\,D_{\mathrm{R}})$ to ensure that the correlation peaks of $\mathcal{F}(I)$ did not overlap, which corresponds to about~$1.5$ interfringes per~$\lambda/D$. We use $1024\times1024$ pixel interferential images with~$4$\,pixels for the smallest interfringe over~$\mathcal{R}$~(Shannon criteria). Compared to the classical sampling used in Guyon's and Codona's devices of $2$~pixels per~$\lambda/D$, the SCC image is $6/2=3$ times oversampled, which reduces the field of view a priori. However, if the read out noise is not a limitation, this oversampling is not a problem since the interesting area~$\mathcal{H}$ is given by the number of actuators of the DM. We consider static aberrations in the instrument upstream of the coronagraph. We adopt a $20$\,nm\,rms amplitude with a spectral power density varying as $f^{-3}$, where $f$ is the spatial frequency, which corresponds to typical VLT optic aberrations~\citep{Borde06}. We simulate an $8$\,m-diameter space telescope with a $50\%$ throughput pointing a G2~star at $10$\,parsec. The quantum efficiency of the detector is $50\%$. We consider photon noise, set the read out noise to $5e-$ per pixel, and consider the zodiacal light to be a uniform background at~$22.5$\,mag.arcsec$^{-2}$. We have not used any linear approximation to simulate the focal plane images. 

By correcting the wavefront errors, we improve the coronographic rejection and the reference intensity~$I_{\mathrm{R}}$ becomes dominant in the science image~$I$~(Eq.~\ref{eq : intens_focal}). We adjust the calibrated neutral density in the reference channel at each step. To determine the value of the neutral density, we estimate the ratio~$r$ of the incoming energies from the image and reference channels in the center of the image to be $r=\int_{\mathcal{H'}}{(I(\alpha)-I_{\mathrm{R}}(\alpha))}/\int_{\mathcal{H'}}{I_{\mathrm{R}}(\alpha)}$ where $\mathcal{H'}$ represents the~$(22\lambda_0/D)^2$ centered on-axis area, and  optimize fringe contrast for the next step. The neutral density transmission is~$1$ at step~$0$ and~$5.6\times 10^{-3}$ at step~$3$. At~$5\,\lambda/D$, the average number of photons per pixel is about~$1.3$ and $50$ for the reference channel, and~$150$ and~$115$ for the image one, respectively at steps~$0$ and $3$. Finally, the intensity in $\mathcal{H}$ decreases as the coronagraphic rejection increases. At each step, we adjust the exposure time to optimize the signal-to-noise ratio in the $16$-bit dynamic range of the detector.

    \begin{figure}[!ht]
      \resizebox{\hsize}{!}{\includegraphics{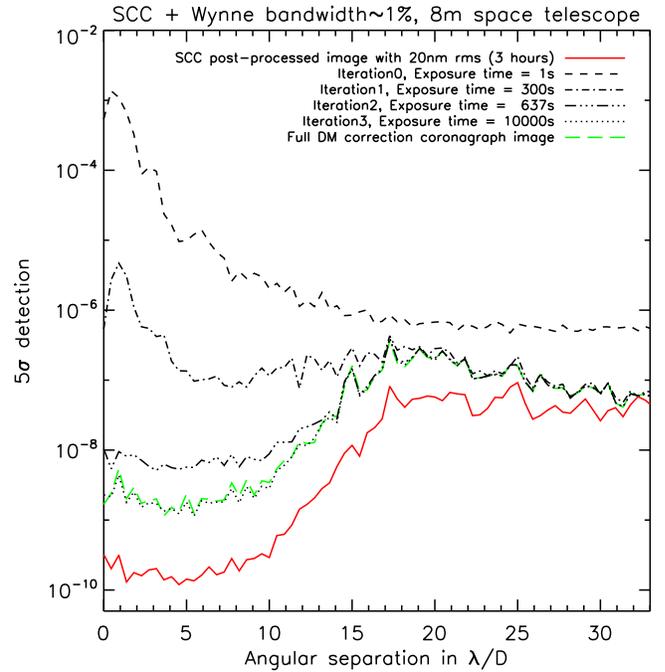}}
      \caption[SCC $5\sigma$ detection limit]{\it $5\sigma$ detection limit vs angular separation.}
      \label{fig : detec}
    \end{figure}
        We define the $5\,\sigma$ detection~$d_{5\,\sigma}$ to be
    \begin{equation}
      d_{5\,\sigma}(\rho) = \frac{5\,\sigma(\rho)}{I_0},
    \end{equation}
where $\sigma(\rho)$ is the azimuthal standard deviation of the considered image at the radial separation of $\rho$ and $I_0$ the maximum intensity of the central star without a coronagraph. We plot~$d_{5\,\sigma}$ for the interferential image~$I$ versus angular separation for several iterations of the correction~(Fig.\,\ref{fig : detec}). The $5\sigma$ detection limit corresponds to an azimuthal average~\citep{Cavarroc06}. In the figure we specify the exposure time of each step. In iteration~$0$, we measure the coronagraphic residue due to the $20$\,nm\,rms static aberrations without any correction. The algorithm converges in a few steps~($\sim 3$). The dashed green line represents the coronagraphic image, without SCC, computed with a full correction by the $32\times32$~DM. This curve is almost surperimposed on the curve of iteration~$3$. This illustrates that the SCC is limited by the aberrations linked to the DM uncorrectable high-order frequencies. The level of this limit depends only on the number of actuators of the DM and the initial aberration level~\citep{Borde06}. To improve the performance, we may increase the number of DM actuators or use higher quality optics. In a second step, we apply to the final iteration image the post-processing algorithm that we presented in previous papers\,\citep{Baudoz06,Galicher07}. The~$5\sigma$ detection limit of the SCC post-processed image is plotted in Fig.\,\ref{fig : detec}~(full red line). The increase in the faintness corresponding to the $5\,\sigma$ detection is about~$10^{5}$ at $5\lambda_0/D$ in a few steps. An Earth-like planet, $2\times10^{-10}$ fainter than its host star, is detected at the $5\,\sigma$ confidence level in about~$3$\,hours. Contrast outside $\mathcal{H}$ is improved slightly during the first steps because both the reference flux~(neutral density) and the corresponding noise decrease.
    
    \begin{figure}[!ht]
      \resizebox{\hsize}{!}{\includegraphics{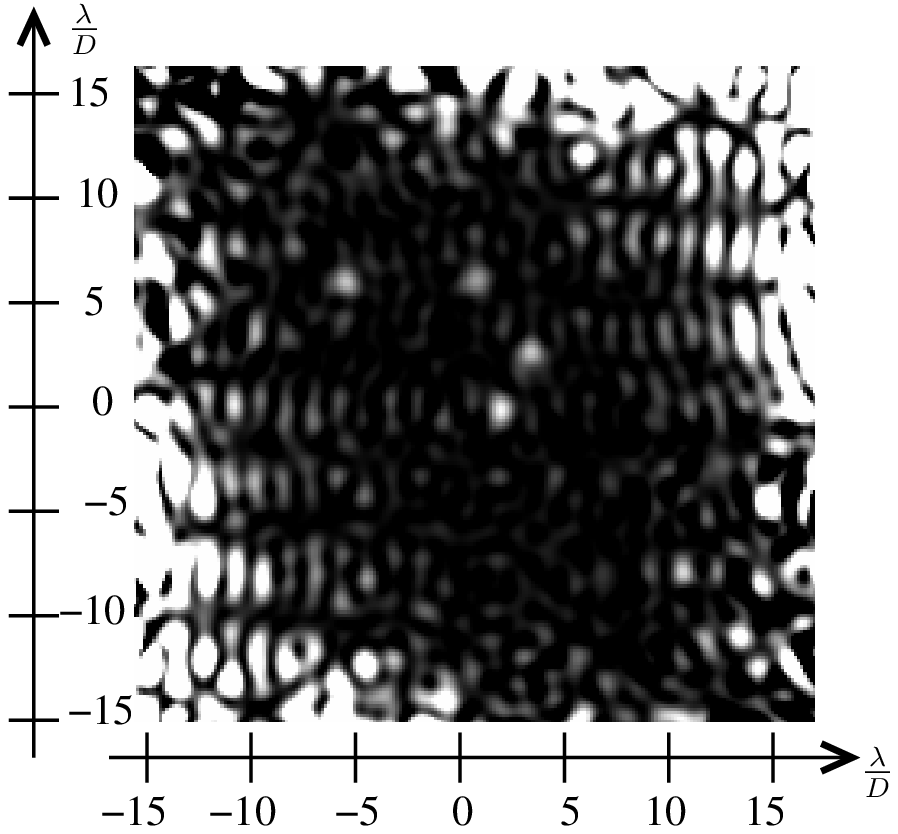}}
      \caption[Companions detection]{\it SCC post-processed image of the final iteration image corresponding to a total exposure time of $\sim 3$\,hours. $2\times10^{-10}$ companions are present at $1$, $3$, $5$, and $7\lambda_0/D$ on a spiral. The field of view is about~$32\times32\,\lambda/D$. The intensity scale is linear.}
      \label{fig : comp}
    \end{figure}
    Under the same assumptions, we simulate four~$2\times10^{-10}$ companions at $1$, $3$, $5$, and $7\lambda_0/D$~($0.02$, $0.06$, $0.10$ and $0.14$\,arcsec), including their photon noise. As shown in Fig.\,\ref{fig : comp}, these Earth-like planets are detected in the SCC post-processed image after a total exposure time $T$ of $\sim 3$\,hours. The accuracy in the measured positions is a fraction of $\lambda/D$~(lower than $\lambda/(2\,D)$ ). Fluxes are determined with a precision better than $20\%$ for the three most off-axis companions. The coronagraph degrades the accuracy of the measured flux of the closest companion~($1\lambda_0/D$): the image of this companion appears to be slightly distorted because the Earth contrast is just above the detection limit~(Fig.~\ref{fig : detec}) and a residual speckle is present at that position. The efficiency of the post-processing algorithm should be improved in future studies. We note that the correction area is larger in the fringe direction~(from top-left to bottom-right) because of the residual chromatic dispersion effect~(Eq.~\ref{eq : I-_0}).

    Similar results for high contrast imaging were demonstrated by~\citet{Trauger07} in a laboratory experiment. They achieved a high contrast of $10^{-9}$ in polychromatic light~($\Delta\lambda\simeq0.02\,\lambda_0$), corresponding to a $5\,\sigma$~detection of $5\times10^{-9}$.
    
    \section{Conclusions}
    We have numerically demonstrated that the SCC associated with a $32 \times 32$ DM  enables us to detect Earths from space in a few hours when using realistic assumptions~(zodiacal light, photon noise, read out noise, VLT pupil aberrations, and $20\%$ bandwidth). SCC could be a good candidate to be implemented in the next generation of space telescopes. The technique involves two steps. We first use SCC to estimate wavefront errors and operate a DM that completes the correction in a few steps. To overcome the limitation linked to the DM uncorrectable high-order frequencies, we apply  to the final iteration image, the SCC post-processing algorithm. This post-processing has yet to be optimized.

SCC is one of the techniques under investigation for the E-ELT planet finder so-called EPICS. For this reason, we propose to consider the impact of different parameters, such as amplitude errors and turbulence residuals on the SCC performance. We will also test the compensation for amplitude errors proposed by~\citet{Borde06}. A preliminary study, which assumes a more realistic coronagraph~(achromatic Four Quadrant Phase Mask), indicates that our algorithm converges but more slowly than with a perfect coronagraph. The quality of the reference beam should not be important for SCC because of the filtering by the pinhole and the reduction in the beam diameter~($D_R$), which induces a wide diffraction pattern in the focal plane. Experimental validations of the SCC technique are also planned soon.
    
    We thank Pascal~Bord\'{e} and Anthony~Boccaletti for useful discussions.

\end{document}